\documentclass[RNAAS]{aastex631}

\usepackage{color}
\usepackage[]{hyperref}
\usepackage{mathptmx}
\usepackage{amsmath}

\begin{document}

\title{UV Slopes  At All Redshifts Are Consistent with $H=1$ Stochastic
Star Formation Histories}
\correspondingauthor{Daniel D.~Kelson}
\email{kelson@carnegiescience.edu}


\author[0000-0003-4727-4327]{Daniel D.~Kelson}
\affiliation{Carnegie Observatories, 813 Santa Barbara Street, Pasadena, CA 91101, USA}

\author[0000-0002-8860-1032]{Louis E.~Abramson}

\affiliation{Carnegie Observatories, 813 Santa Barbara Street, Pasadena, CA 91101, USA}

\begin{abstract}
\noindent Multiple investigations support describing galaxy growth as a stochastic process with correlations over a range of timescales governed by a parameter, $H$, empirically and theoretically constrained to be near unity.
Here, we show that the distribution of UV-slopes, $\beta$, derived from an ensemble of theoretical $H=1$ star formation histories (SFHs) is consistent with data at all redshifts $z\le 16$.
At $z=0$, the median value $\langle\beta_{H=1}\rangle=-2.27$ agrees well with the canonical $\beta_0=-2.23$ for local starbursts \citep{meurer1999}.
At $4\lesssim z\lesssim16$, JWST data span the model distribution's 2nd to 98th percentiles.
Values of $-2.8\le \beta \le -2.5$ should  be common in early galaxies without reference to exotic stellar populations---arising solely from a null hypothesis of $H=1$ for the underlying diversity of galaxy growth histories.
Future data should be interpreted with this fact in mind.
\end{abstract}




\section{Introduction} 

Restframe ultraviolet light is a proxy for ongoing star formation \citep[e.g.,][]{calzetti1994}. Observations of local starbursts 
suggest that vigorous star formation today leads to an unreddened UV spectral slope of $\beta_0 = -2.23$ \citep{meurer1999}.
At higher redshift, the distribution of $\beta$ may shift to more negative values, or bluer UV continua \citep[e.g.,][]{bouwens2014}.
Such trends might reveal changes in the properties of stellar populations with lookback time, with some theoretical work indicating that $\beta\ll -2.5$ suggests the presence of Population III stars \citep{robertson2010}. Such extreme negative values have now been observed at high redshifts with JWST \citep[e.g.,][]{nanyakkara2022}.

However, $\beta$ is sensitive not only to the properties of the underlying stellar population, but also a galaxy's star formation history (SFH).
Physical interpretations that invoke exotic physics must therefore account for the true diversity of SFHs based on a null hypothesis derived from a realistic suite of growth trajectories.




Stochastic processes provide a useful framework for generating such trajectories \citep{kelson2014}.
Cooling, heating, turbulence, the formation of star clusters---and any resulting feedback---correlate changes to galaxy star formation rates (SFRs) over a broad range of timescales, allowing one to express stellar mass growth with a relatively simple parameterization of 
these spectrum of correlated timescales, $H$---the Hurst parameter \citep[e.g.,][]{mandelbrot1968}.
From a theoretical perspective, gravitational collapse itself drives systems to $H=1$  \citep{kelson2020}, implying that recent changes to accretion rates---that serve to fuel star formation---are maximally correlated with all previous changes that have occured.

Ensembles of theoretical SFHs assuming $H=1$ have been shown to reproduce distributions of specific star formation rates and restframe colors \citep[e.g.,][]{kelson2014}, generate the stellar mass function \citep{kelson2016},
and match the passive galaxy fraction at appropriate masses and/or epochs \citep{abramson2020}. Here, we show that they also reproduce measurements of $\beta$ at $z\lesssim16$.




\begin{figure}[h!]
\centering
\includegraphics[width=6.0in]{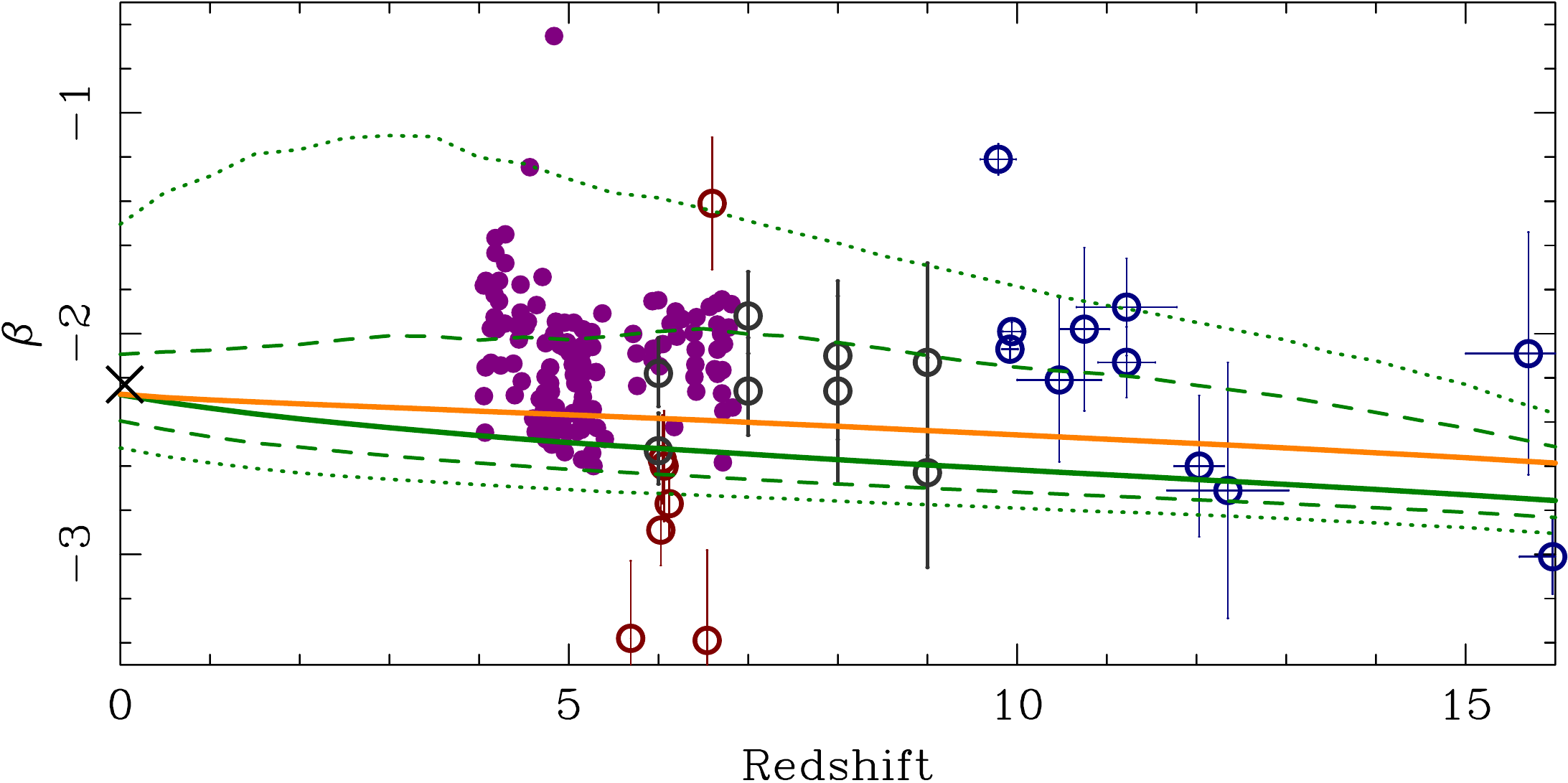}
\caption{Green lines show the distribution of unattenuated UV-slopes as a function of redshift assuming galaxy star formation is a stochastic process that starts at $z=20$ with $H=1$.
These models also assume $\rm [Fe/H]=-1.2$ dex for the metallicity of the stars.
The solid line shows the median $\beta$ value with redshift; dashed lines follow the 10th and 90th percentiles; dotted lines show the 2nd and 98th.
The black X marks $\beta_0=-2.23$ from \citet{meurer1999}---the unattenuated value derived from local starbursts. Data from \citet[red]{jiang2020}, \citet[black]{bhatawdekar2021}, \citet[purple]{nanyakkara2022}, and \citet[blue]{atek2022} fall 
within the predicted range at all epochs. Any reddening would skew $\beta$ to the positive direction. For comparison, the orange line traces the median $\beta(z)$ for solar metallicity.}
\label{fig:evol}
\end{figure}


\section{Expectations for Unattenuated UV Slopes} 

Figure \ref{fig:evol} shows the $H=1$ stochastic expectation for the distribution of $\beta$ for galaxies at $z\le 16$ with ongoing star formation. Model SFRs were computed over a 10 Myr timeframe. Model galaxies were classified as ``star forming'' if they had sSFRs above a threshold placed $4\sigma$ below the median (results are insensitive to these details). We used the Flexible Stellar Population Synthesis library \citep{conroy2010} assuming a \citet{chabrier2003} IMF, 
$\rm [Fe/H]=-1.2$, no attenuation/reddening, and an onset of star formation at $z=20$. We compute $\beta$ using the standard \citet{calzetti1994} bandpasses. A sample of $4\,000$ model histories defines the distribution at all redshifts.

The solid green line shows the predicted evolution of the median unattenuated UV slope, $\beta$, with redshift.
At $z=0$, this yields  $\langle\beta_{H=1}\rangle=-2.27$, in good agreement with \citet{meurer1999}'s observation of $\beta_0=-2.23$.\footnote{We find $\langle\beta_{H=1}\rangle=-2.24$ at $z=0$ when assuming $\rm [Fe/H]-2.4$ dex
(not shown).}
At higher redshift, the data span the predicted 2nd to 98th percentiles (dotted lines) with the vast majority of $4\lesssim z \lesssim 12$ measurements lying
inside the 10th to 90th (dashed lines). 

The shift of the $\beta$ distribution to bluer slopes at higher redshifts in this plot {\it exclusively\/} reflects the changing contrast between young and 
(relatively) old stars. The evolution of $\beta$ is not as strong when using solar stellar matallicities (orange solid line in Figure \ref{fig:evol}). 
Due to the diversity of recent star formation activity in the $H=1$ SFHs, the distribution of expected $\beta$ values 
is substantially asymmetric/non-Gaussian, even absent any dust or metallicity effects. Conversely, high-redshift $A_V$ and dust content estimates should be performed with respect to a $\beta(z)$ locus defined by these model curves, or similarly generated ones using the best available stellar metallicities.

High-redshift restframe UV continuum observations may provide insights into novel phenomena, including the first stars.
However, measurements of $\beta$ at $z<16$ can so far be explained by invoking nothing beyond a broad diversity of star formation histories.
Specifically, these data---and many others---are consistent with galaxies growing as $H=1$ stochastic processes.

With $H=1$ SFHs providing a prior on the underlying growth trajectories of early galaxies, values of $-2.8\le \beta \le -2.5$ should be common \citep[e.g.,][]{atek2022} irrespective of the presence of, e.g., Population III stars. Towards lower redshifts, the wider range and higher median is expected. Should star 
formation begin at $z<20$ for some subsamples of galaxies---perhaps due to differences in environmental density---more extreme negative values of  $\beta$ are possible.

Many salient aspects of galaxy evolution can be interpreted through the lens of stochastic SFHs. Accurate analyses 
of high redshift data---especially claims of new physics---must account for the impact and implications of this more complete span of growth 
trajectories compared to typical (and typically parametric) models. Fortunately, $H=1$ SFHs are easy to generate, making such null hypotheses 
easy to test. A sample of $100\,000$ such histories, with $10\,000$ timesteps can be downloaded from
\dataset[https://code.obs.carnegiescience.edu/Stochastic]{https://code.obs.carnegiescience.edu/Stochastic}.




\begin{acknowledgements}
We thank Dr.~A.~ J.~Benson for thoughtful comments.
\end{acknowledgements}


\clearpage


\begin{thebibliography}{}

\expandafter\ifx\csname natexlab\endcsname\relax\def\natexlab#1{#1}\fi

\bibitem[Abramson \& Kelson(2020)]{abramson2020} Abramson, L.~E. \& Kelson, D.~D.\ 2020, RNAAS, 4, 236

\bibitem[Atek et al.(2022)]{atek2022} Atek, H., Shuntov, M., Furtak, L.~J., et al.\ 2022, arXiv:2207.12338

\bibitem[Bhatawdekar \& Conselice(2021)]{bhatawdekar2021} Bhatawdekar, R. \& Conselice, C.~J.\ 2021, \apj, 909, 144

\bibitem[Bouwens et al.(2014)]{bouwens2014} Bouwens, R.~J., Illingworth, G.~D., Oesch, P.~A., et al.\ 2014, \apj, 793, 115

\bibitem[Calzetti et al.(1994)]{calzetti1994} Calzetti, D., Kinney, A.~L.,  \& Storchi-Bergmann, T. 1994, \apj, 429, 582,

\bibitem[Chabrier(2003)]{chabrier2003}Chabrier, G. 2003, \pasp, 115, 763.

\bibitem[Conroy \& Gunn(2010)]{conroy2010} Conroy, C., \& Gunn, J. E. 2010, \apj, 712, 833,

\bibitem[Jiang et al.(2020)]{jiang2020} Jiang, L., Cohen, S.~H., Windhorst, R.~A., et al.\ 2020, \apj, 889, 90

\bibitem[Kelson(2014)]{kelson2014} Kelson, D.~D.\ 2014, arXiv:1406.5191

\bibitem[Kelson et al.(2016)]{kelson2016} Kelson, D.~D., Benson, A.~J., \& Abramson, L.~E.\ 2016, arXiv:1610.06566

\bibitem[Kelson et al.(2020)]{kelson2020} Kelson, D.~D., Abramson, L.~E., Benson, A.~J., et al.\ 2020, \mnras, 494, 2628

\bibitem[Mandelbrot \& van Ness(1968)]{mandelbrot1968} Mandelbrot, B. B., \& van Ness, J. W. 1968, SIAM Review, 10, 422


\bibitem[Meurer et al.(1999)]{meurer1999} Meurer G.~R.,  Heckman T.~M.,  \& Calzetti D.,\ 1999, \apj, 521, 64

\bibitem[Nanayakkara et al.(2022)]{nanyakkara2022} Nanayakkara, T., Glazebrook, K., Jacobs, C., et al.\ 2022, arXiv:2207.13860


\bibitem[Robertson et al.(2010)]{robertson2010} Robertson, B.~E., Ellis, R.~S., Dunlop, J.~S., et al.\ 2010, \nat, 468, 49

 
 

\end{thebibliography}
\end{document}